\documentclass[aps,amsmath,amssymb,prl,showpacs]{revtex4}
\usepackage{bm}
\usepackage{epsfig}
\usepackage[dvips]{color}

\newcommand{\tav}[1]{\left\langle #1\right\rangle} 

\begin{document}

\title{$1/f$ Noise in a one-dimensional charge density wave system}

\author{Zbigniew R. Struzik}
\affiliation{Educational Physiology Laboratory, Graduate School of
  Education, The University of Tokyo, 7--3--1 Hongo, Bunkyo-ku, Tokyo
  113--0033, Japan}
\affiliation{PRESTO, Japan Science and Technology Agency,
Kawaguchi, Saitama 332--0012, Japan}

\author{Andreas Glatz}
\affiliation{Institut f\"ur Theoretische Physik, Universit\"at zu
K\"oln, Z\"ulpicher Str. 77, 50937 K\"oln, Germany}

\author{Mai Suan Li}
\affiliation{Institute of Physics, Al. Lotnikow 32/46, 02-668
Warsaw, Poland}

\begin{abstract}
The current noise in a classical one-dimensional charge density
wave system is studied in the weak pinning regime by solving the
overdamped equation of motion numerically. At low temperatures and
just above the zero temperature depinning threshold, the power
spectrum of the current noise $S(f)$ was found to scale with
frequency $f$ as $S(f) \sim f ^{-\gamma}$, where $\gamma \approx
1$, suggesting the existence of {\it flicker noise}. Our result is
in agreement with experimental findings for quasi-one-dimensional
charge density wave systems and provides the first evidence of
$1/f$ behavior obtained from first principles.
\end{abstract}


\maketitle

\section{Introduction}

The noise in any system is characterized in terms of the shape of
its spectral power density $S(f)$, which may be measured directly
in experiments.  If $S(f) \sim f^{-\gamma}$ and $\gamma \approx
1$, then such a noise is referred to as $1/f$ or flicker noise.
Its existence has attracted the attention of researchers from
various branches of the natural sciences for many years. As for
physical systems, $1/f$ noise has been observed as fluctuations in
the currents of diodes, vacuum tubes and transistors, the
resistance of carbon microphones, metallic thin films and
semiconductors~\cite{Voss79,Hooge76,Bochkov83}, the magnetization
in spin glasses~\cite{Reim86} and in many other systems.

Despite a substantial amount of experimental and theoretical
effort, a unique underlying mechanism for the flicker noise
remains unknown and still poses an open question as to its
universal origins~\cite{Milotti01, Sethna01, Antal01}. However,
there is now increasing evidence that the non-trivial exponent, at
least in electronic systems, may come from equilibrium dynamics
in the presence of disorder~\cite{Kogan96} when energy barriers exceed
the typical thermal energy.

$1/f^{\gamma}$ noise in charge density wave (CDW)
systems~\cite{Gruner88} has been studied mainly in NbSe$_3$ and
TaS$_3$ materials \cite{Kogan96}. The first experiment was done on
a bulk NbSe$_3$ sample by Richard {\em et al.} \cite{Richard82}
who found $\gamma \sim 0.8$. Studying transport properties of the
quasi-one-dimensional CDW material TaS$_3$ at low temperatures,
Zaitsev--Zotov~\cite{Zaitsev93} observed that slightly above the
depinning threshold of the driving electric field, the exponent
$\gamma$ for the current noise is equal to $\gamma \approx 1.2$.

It should be noted that a phenomenological model based on
fluctuations in impurity pinning force due to deformations of the
sliding condensate was proposed to explain the broad band $1/f$
noise in CDW systems \cite{Bhattacharya85}. However, neither
theoretical nor numerical estimation of $\gamma$ has been provided
so far. The aim of the present paper is to compute $\gamma$ from
{\em first principles} with the help of a one-dimensional
classical model for CDWs~\cite{Gorkov77}. The current was obtained
through numerical simulation of the overdamped equation of motion.
The $1/f$ scaling is evaluated using the so-called {\it Wavelet
Transform Modulus Maxima} (WTMM) method~\cite{Arneodo}. The
exponent $\gamma$ was found to depend on $T$. At low temperatures
($T \leq 0.1$), in agreement with the
experiments~\cite{Zaitsev93}, we obtain $\gamma \approx 1.2$ in
the crossover regime. Exponent $\gamma$ drops with increasing $T$
and the "exact" $1/f$-noise is observed at $T \approx 0.3$ where
$\gamma$ becomes 1. This interesting result is indicative of the
possible occurrence of $1/f$ noise. At high temperatures $\gamma$
takes on the white noise value 0.

Notably, the observed $\gamma \approx 1$ is not related to the
second order depinning transition behavior at $T=0$. Due to the
asymptotic uniqueness of the sliding state \cite{Middleton92},
this critical point dynamics scenario leads to the `trivial'
exponent $\gamma \approx 2$~\cite{Narayan92,Fisher98}.
Additionally, the observed `flicker' noise behavior $\gamma
\approx 1$ gains on its scaling range with increased distance to
the critical point of the second order depinning transition.

Based on unusual current--voltage
characteristics~\cite{Zaitsev93,Zaitsev94,Zaitsev97a,Zaitsev97b},
Zaitsev-Zotov suggested that at low temperatures the quantum creep
dynamics may play an important role and proposed the crossover
from classical to quantum creep regime as an alternative
explanation for experimental results. The strength of the quantum
fluctuations in 1D CDW systems can be estimated by a dimensionless
parameter $K$~\cite{GlaNa02} which is proportional to
$\sqrt{m^*/m}$, where $m^*$ is the effective band mass. This
quantity is of the order $10^{-2}$ to
$10^{-1}$~\cite{Maki95,Gruner88}, indicating irrelevance (to 1/f
noise) of quantum effects at low temperatures. Furthermore, our
simulation results on transport properties~\cite{Glatz02} also
suggest, in comparison to experiments, that quantum fluctuations
do not have any visible effect with respect to the strength of the
driving forces under consideration (see also discussion
in~\cite{Kiss88,Handel88}). On the other hand, due to the small
parameter $K$ the core action of phase slips in the bulk is large
($S_{\textit{core}}\propto 1/K$)~\cite{GlaNa02} and hence the
probability of phase slips which are proportional to
$e^{-S_{\textit{core}}}$ becomes small. It decreases even more
under a renormalization group transformation, such that we can
neglect phase slips in our simulations. Therefore we will use the
one-dimensional classical model without phase slips to study the
current noise in CDW systems.

\section{Model}

The charge--density $\rho(x)$ of an 1D CDW can be expressed as
$\rho(x)=\rho_0(1+Q^{-1}\partial_x\varphi(x))+\rho_1\cos{\big(Qx+\varphi(x)\big)}$,
where $Q=2k_F$ denotes the wave vector of the undistorted wave,
$k_F$ the Fermi wave vector and $\varphi(x)$ a slowly varying
phase variable. $\rho_0=Q/\pi$ is the mean electron density and
$\rho_1$ is proportional to the amplitude of the complex order
parameter~\cite{GlaNa02}. The Hamiltonian of the phase field is
then given by
\begin{equation}
{\cal H}=\int dx\,\Bigg\{\frac{c}{2}
\left(\frac{\partial}{\partial x}\varphi\right)^2-\sum\limits_i
V_i\delta(x-x_i)\times\rho_1\cos{\big(Qx+\varphi(x)\big)} +  E x
\partial_x\varphi(x) \Bigg\}\,,\label{fullHamiltonian}
\end{equation}
where $c=\frac{\hbar v_F}{2\pi}$ is the elastic constant with the
Fermi--velocity $v_F$, and $V_i$ and $x_i$ denote the strength and
the position of the impurity potential acting on the CDW,
respectively; $E$ is the external electric field or driving force.

Our numerical studies are done in the weak pinning limit, i.e.
when the Fukuyama--Lee length~\cite{FuLee78} $L_c=(c/V)^{2/3}$ is
large compared to the mean impurity distance $l_{\textit{imp}}$.
Therefore we will restrict ourselves in the following to the case
$L_c\gg l_{\textit{imp}}\gg Q^{-1}$, where the full Hamiltonian
(\ref{fullHamiltonian}) can be reduced to a random field
XY--model:
\begin{equation}
   {\cal H}=\int dx\left\{\frac{c}{2}
   \left(\frac{\partial}{\partial x}\varphi\right)^2\right.
  \left.-V\cos{\big(\varphi-\alpha(x)\big)} \, - \,
   E \varphi(x)\right\}\, .
   \label{Hamiltonian}
\end{equation}
Here $\alpha(x)$ is a random phase with zero average and
$\overline{e^{i\big(\alpha(x)-\alpha(x^{\prime})\big)}}=
\delta(x-x^{\prime})$, where the overbar denotes the averaging
over disorder realizations. $V$ is defined by
$\overline{(V_i\rho_1)(V_j\rho_1)}\equiv V^2\delta_{ij}$
($\overline{V_i}=0$). The equation of motion of the (overdamped)
CDW is given by a {\it Langevin equation}
\begin{equation}
   \frac{\partial\varphi}{\partial t}=-\gamma
   \frac{\delta {\cal H}}{\delta\varphi}
   +\eta(x,t)\,,
   \label{Langevin}
\end{equation}
where $\gamma$ is a kinetic coefficient and $\eta(x,t)$ a
Gaussian thermal noise characterized by $\big<\eta\big>=0$ and
$\big<\eta(x,t)\,\eta(x^{\prime},t^{\prime})\big>=
2T\gamma\,\delta(x-x^{\prime})\,\delta(t-t^{\prime})$.

The length scale $L_c$ sets an energy scale
$T^{\ast}=\left(c^{\phantom{1}}V^2\right)^{1/3}=c^{\phantom{1}}L_c^{-1}$.
We will rescale time by $L_c/\gamma T^{\ast}$, temperature by
$T^{\ast}$, and the external field $E$ by $E^{\ast}$, where
$E^{\ast}=T^{\ast}/L_c$ is of the order of the $T=0$ depinning
threshold field $E_c$. In the following, $E$ denotes the
rescaled and dimensionless quantity.

Solving the discretized version of Eq. \ref{Langevin} one can find he time dependent
current $j_{\textit{cdw}}(t)$ which is defined as \cite{Glatz01}
\begin{equation}
j_{\textit{cdw}}(t)=\overline{\tav{\frac{\partial\varphi(x,t)}{\partial
t}}_x}, 
\end{equation}
where $\tav{...}_x$ denotes the average over positions. 

\section{Simulation}

The effect of disorder on the dynamical behavior of the
one-dimensional charge density wave model (\ref{Hamiltonian}) at
low temperatures was studied~\cite{Glatz01} with the help of the
discretized version of equation (\ref{Langevin}) and it was found
that, contrary to high dimensional systems~\cite{Natter90}, the
dependency of the creep velocity on the electric field is
described by an analytic function. The current noise spectrum was
not, however, explored. Following Ref. \cite{Glatz01}, the
equation of motion (\ref{Langevin}) is integrated by a modified
Runge--Kutta algorithm suitable for stochastic
systems 
with periodic boundary conditions.

\begin{figure}[htb]
 \includegraphics[width=1.0\linewidth]{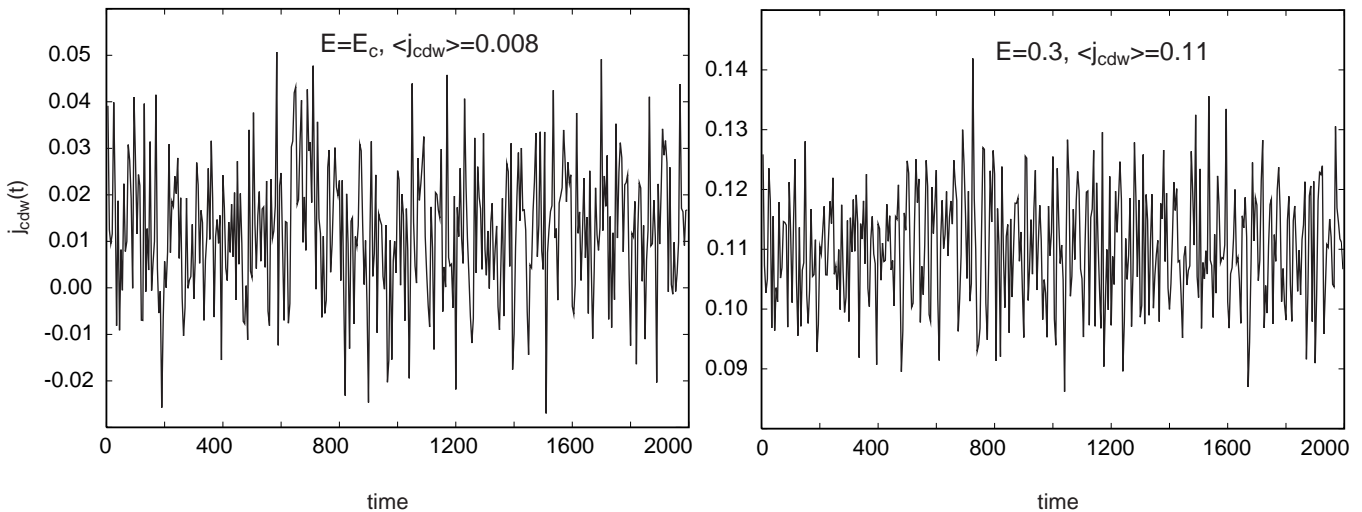}
\caption{Typical time dependence of the CDW current for
$E=E_c=0.22$ (left panel) and $E=0.3$ (right panel) at $T=0.1$ for
one disorder realization. The time and disorder averaged values of
$j_{\textit{cdw}}(t)$ are shown next to the curves ($\langle
j_{\textit{cdw}}\rangle$). We took $N=5000$ and the results are
averaged over 1000  and 500 samples for $E=E_c$ and $E=0.3$,
respectively. An initial (dimensionless) time interval of length
$\approx 2000$ is discarded, such that the system is in the steady
state at time $0$.}\label{fig1}
\end{figure}

Throughout this paper, we use a system size of $N=5000$ and
average the results over typically  $N_s= 1000$ disorder
realizations. Larger system sizes do not change the results
substantially.
Fig. \ref{fig1} shows  the typical time evolution of
$j_{\textit{cdw}}(t)$ for $E=E_c = 0.22$ \cite{Glatz01} (upper
panel) and $E=0.3$ (lower panel) at temperature $T=0.1$. One can
see that the current exhibits strong fluctuations. The time
averaged values are $\langle j_{\textit{cdw}}\rangle=0.008\pm
0.003$ and $\langle j_{\textit{cdw}}\rangle=0.112\pm 0.006$ for
$E=E_c$ and $E=0.3$, respectively. The spike structure is also
seen, but less pronounced compared to the experimental
data~\cite{Zaitsev93}. Nevertheless, the patterns for the two
values of $E$ look similar.

Zaitsev--Zotov studied~\cite{Zaitsev93} the current noise spectrum
for applied electric fields with averaged driving current
$\tav{I}\geq 220$pA. Using Fig. 1 from Ref. \cite{Zaitsev93} one
can see that the threshold electric field  in these experiments is
$E_c \approx 35$V/cm and the averaged currents of $\tav{I}=220$pA
and $\tav{I}=2.4$nA at $T=2.4$K correspond to electric fields
$E\approx 40$V/cm and $E\approx 50$V/cm, respectively, i.e. the
electric fields used are greater than the threshold field.
Therefore we will restrict our spectrum analysis to  $E\geq E_c$.

\section{Scaling exponent estimation}

It should be noted that in the case of non-stationary behavior of
the CDW current as in our simulations, the standard Fourier
transformation is not suitable for determining the exponent $\gamma$
and one should, therefore, employ more sophisticated methods. We
have chosen the WTMM method~\cite{Arneodo} for its superior
properties in non-parametric scaling exponent
estimation~\cite{Audit02} in the presence of polynomial
non-stationarities.
In particular, attempts
to reduce the non-stationary behavior of the current by discarding an
initial time interval (as in~\ref{fig1}) cannot generally guarantee
reaching a steady state,
since the relaxation time to a steady state can be very
long (see the remark in Ref.~\cite{Middleton93}).

The ability of the wavelet transform to provide unbiased scaling
estimates of non-stationary signals is due to the property of
orthogonality to polynomials up to the degree $n$ of the base
functions, of the so-called analyzing wavelets $\psi$ with $m$
`vanishing moments':
\[\int_{-\infty}^{+\infty}x^n\,\psi(x)\,dx=0\hspace{0.2in}\forall n,\;0\leq n <m\;.\]
The transform is defined as the inner product of the function $f(x)$
and the dilated and translated wavelet $\psi(x)$:

\begin{equation}
(Wf)(s,b)=\frac{1}{s}\;\int dx
\;f(x)\;\psi(\frac{x-b}{s})\;,
\label{eqn:WT}
\end{equation}
where $s,b \in \mathbb{R}$ and $s>0$ for the continuous version
(CWT), which among other properties ensures local blindness to the
polynomial bias. Indeed, the wavelet transform decomposes the
signal into scale (and thus frequency) dependent components (scale
and position localized wavelets), comparable to frequency
localized sines and cosines based Fourier decomposition, but with
added position localization. This localization in both space and
frequency, together with the wavelet's orthogonality to polynomial
bias, makes it possible to access even weak scaling behavior of
singularities $h(x_0)$, otherwise masked by the stronger
polynomial components:
\[
f(x)_{x_0} = c_0 + c_1(x-x_0) + \dots +c_n(x-x_0)^m + C|x-x_0|^{h(x_0)}\;,
\] where function $f$ is represented through its Taylor expansion around $x=x_0$.

In the generic multifractal formulation of the WTMM
formalism~\cite{Arneodo}, the moments~$q$ of the measure distributed
on the WTMM tree are taken to obtain the dependency of the scaling
function $\tau(q)$ on the moments~$q$:

\[{\cal Z}(s,q) \sim s^{\tau(q)},\]
\noindent
where ${\cal Z}(s,q)$ is the partition function of the $q$-th
moment of the measure distributed over the wavelet transform
maxima at the scale $s$ considered:

\begin{equation}
{{\cal Z}(s,q) = \sum_{{\Omega}(s)} (Wf\omega_i(s))^q \;,}
\label{eqn:partition_function}
\end{equation}

\noindent
with ${\Omega}(s)= \{ \omega_i(s)\}$ as the set of maxima
$\omega_i(s)$ at the scale $s$ of the continuous wavelet transform
$Wf(s,t)$ of the function $f(t)$, in our case the CDW current;
$f(t)=j_{\textit{cdw}}(t)$.

In particular, scaling analysis with WTMM is capable of revealing
the modal exponent $h(q=0)$ for which the spectrum reaches maximum
value; this $h(q=0)$ corresponds to the Hurst exponent $H$ in the
case of monofractal noise. This exponent is directly linked to the
power spectrum exponent of the (stationary) fluctuations of the
analyzed signal by:  $\gamma=2H+1$, the relation which links the
spectral exponent $\gamma$ with the Hurst exponent $H$.

\begin{figure}[t]
\includegraphics[width=1.0\linewidth]{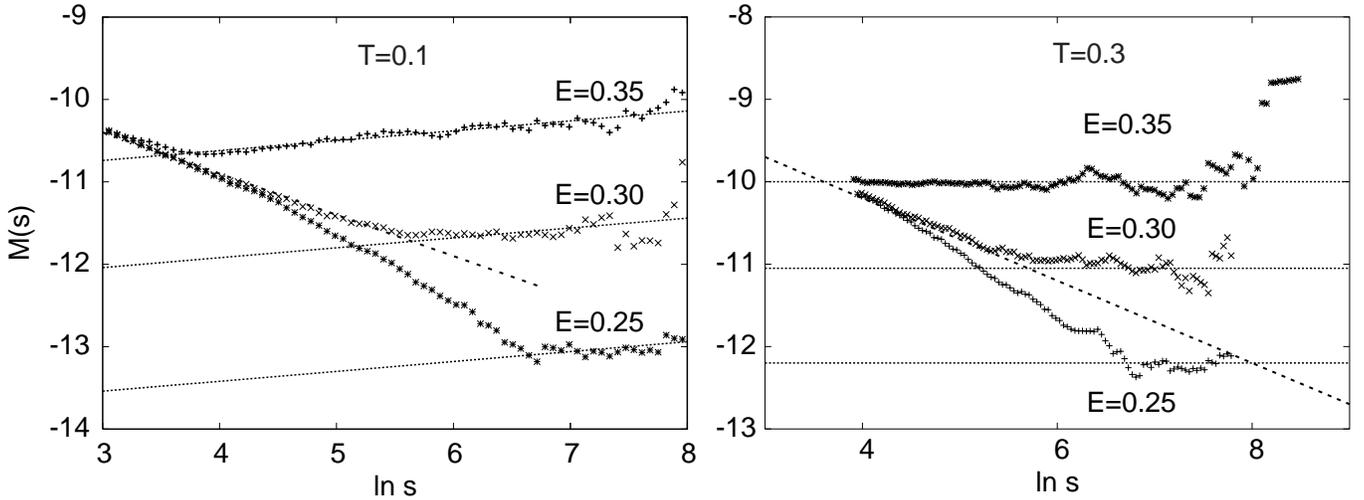}
\caption{Left: {\cal M}(s) versus $\ln s$ for the CDW current
$j_{\textit{cdw}}(t)$ averaged over $N_s=1000$ disorder
realizations and for three values of the external electric field:
$E=0.25$, 0.3 and 0.35 which are higher than $E_c \approx 0.22$.
The dotted straight lines denote the reference slope corresponding
to $\gamma~=~1.2$. The dashed line has the slope $-0.5$, which
corresponds to the flat power spectrum of white noise $\gamma=0$.
Right: The same but for $T = 0.3$. The dotted lines denote the
reference slope $H = 0$ corresponding to $\gamma~=~1.0$. Again,
the dashed line corresponds to white noise.} \label{fig_MF_C}
\end{figure}

In Fig. \ref{fig_MF_C}, the modal scaling exponent has been obtained by a
linear fit over an appropriate scaling range
 from a suitably defined, weighted measure ${\cal M}(s)$ on the WTMM: 

\begin{equation}
h(q=0)=\left.\frac{d\tau(q)}{dq}\right|_{q=0} = \lim_{s\rightarrow
0} \; \frac{{\cal M}(s)}{\log(s)}
\end{equation}
with
\begin{equation}
{{\cal M}(s)= \frac{\sum_{{\Omega}(s)} \log(Wf\omega_i(s)) \;,}
{{\cal Z}(s,0)}}\;, \label{eqn:mean_partition}
\end{equation}

\noindent and for three electric field  values $E$=0.25, 0.3 and
0.35, and for the number of the disorder averaging ensemble fixed
to $N_s=1000$. Consistent with the experimental findings
\cite{Zaitsev93}, the flicker noise region becomes narrower with
decreasing $E$. More importantly, we obtain $\gamma \approx 1.2$
as observed in experiments \cite{Zaitsev93}. Asymptotic transition
to the scaling regime characteristic to uncorrelated behavior
(white noise, i.e., $\gamma=0$) can be clearly identified for all
the values of $E$ shown (see dashed line in Fig. \ref{fig_MF_C}). 

\begin{figure}[]
\includegraphics[width=0.6\linewidth]{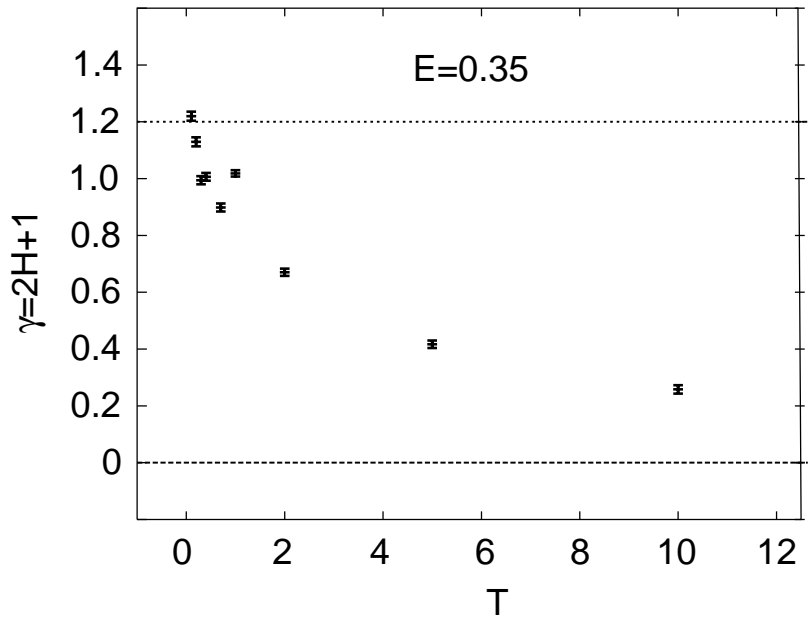}
\caption{The dependence of the exponent $h(q=0)$ on temperature.
Note
  the convergence towards $H=0.1$ corresponding with $\gamma=1.2$
  spectral exponent for $0$ values of temperature and current.}
\label{H_T}
\end{figure}

Fig. 
 \ref{fig_MF_C}) (right)
shows {\cal M}(s) versus $\ln s$ for
$j_{\textit{cdw}}(t)$ for three values of the external electric
field $E=0.25$, 0.3 and 0.35 and $T=0.3$.
Our fitting gives $\gamma =1$, which is
important from the point of view of the exact definition of
$1/f$-noise. In Fig. \ref{H_T}, we provide the dependence of the
exponent $\gamma$ on temperature. Note the convergence towards
$\gamma=1.2$ as the temperature (and the averaged current) approaches
0. The exponent $\gamma$ decays quickly with temperature and we
have the uncorrelated noise value $\gamma = 0$ at high $T$.

The primary question remaining is that of the origins of $1/f$ noise in
the CDW system.
In our opinion, the disorder causes the rugged energy landscape
(similar to the spin glass case) leading to a wide spectrum of
relaxation times. The average over such a spectrum would give rise
to the flicker noise~\cite{Bochkov83}. Our results shown in Fig. \ref{H_T}
support this point of view. Namely, at low temperatures the roughness
of the energy landscape becomes more important and consequently the flicker-like
regime occurs.
Another qualitative scenario~\cite{Milotti01} for the appearance of
the flicker noise in our system is that the CDW may be viewed as
a single particle in a quasi-periodic potential with troughs of
variable depths. Such a simplified model closely resembles the
``many-pendula'' model of the self-organized
criticality~\cite{Bak87} in which the $1/f$ noise should occur.

\section{Conclusion}

In conclusion, using the classical one-dimensional CDW model and
multifractal analysis, we have reproduced the experimental
results on the current noise spectrum. Our simulations support the
existence of $1/f$-noise in this system. To the best of our
knowledge, this is the first evidence of $1/f$ scaling obtained
from first principle based simulation in a physical (i.e. CDW) system.

It would be interesting to check if
a three-dimensional version of our model gives $\gamma \approx 0.8$ obtained
for the bulk NbSe$_3$ sample \cite{Richard82} or if other models should
be implemented to reproduce this experimental result.
The effect of higher dimensionalities, phase slips and quantum fluctuations
on the flicker noise remains a challenge for future studies.

\section{Acknowledgement}

The authors wish to thank A. Ausloos, J. Kertesz, T. Nattermann,
S. Scheidl and S. Zaitsev--Zotov for useful discussions.
AG acknowledges financial support from the Deutsche
Forschungsgemeinschaft through Sonderforschungsbereich 608 and the
Deutscher Akademischer Austauschdienst and
MS Li from the Polish agency KBN.

\end{document}